\documentclass[twoside]{article}
\usepackage[accepted]{aistats_stripped}
\usepackage{amsmath,graphicx}

\usepackage{booktabs}
\usepackage{soul} 
\usepackage{amsmath,amsfonts,amssymb,amsbsy}
\usepackage{url}
\usepackage{xcolor}
\usepackage{footnote}
\usepackage{graphicx}
\usepackage{caption}
\usepackage{multirow}
\usepackage{subcaption}
\usepackage{tikz}
\usepackage{placeins}
\usepackage[titletoc]{appendix}
\usepackage{algorithm}
\usepackage{algpseudocode}

\usepackage[colorlinks,linkcolor=black,citecolor=black,urlcolor=black,filecolor=blue]{hyperref}

\usepackage{ifthen}
\usepackage{tikz,pgfplots}
\usetikzlibrary{matrix}
\usetikzlibrary{calc}

\def\figpdfdir{./} %
\def\figtikzdir{./} %

\newcommand{\minput}[2][]{
\ifthenelse{\equal{#1}{pdf}}
	{ \includegraphics{\figpdfdir #2} }
	{ \tikzset{external/remake next} \tikzsetnextfilename{#2} \input{\figtikzdir #2} }
}

\usetikzlibrary{external}
\tikzset{external/system call={lualatex
	\tikzexternalcheckshellescape -halt-on-error -interaction=batchmode
	-jobname "\image" "\texsource"}}

\usepackage{natbib}
\usepackage{microtype}

\begin{document}
\newcommand{\subjto}{ \, \tx{s.t.} }
\newcommand{\tr}{^{\text{T}} }
\newcommand{\given}{ \,\middle| \,}
\newcommand{\cov}[1]{\text{cov}\left( {#1} \right)}

\twocolumn[

\aistatstitle{Automatic monotonicity detection for Gaussian Processes}

\aistatsauthor{ Eero Siivola \And Juho Piironen \And Aki Vehtari}
\aistatsaddress{ {\tt eero.siivola@aalto.fi} \\ Aalto Univeristy \And {\tt juho.piironen@aalto.fi} \\ Aalto University \And {\tt aki.vehtari@aalto.fi} \\ Aalto University } ]

\begin{abstract}

We propose a new method for automatically detecting monotonic input-output relationships from data using Gaussian Process (GP) models with virtual derivative observations. Our results on synthetic and real datasets show that the proposed method detects monotonic directions from input spaces with high accuracy. We expect the method to be useful especially for improving explainability of the models and improving the accuracy of regression and classification tasks, especially near the edges of the data or when extrapolating.
\end{abstract}

\section{INTRODUCTION}
\label{sec:intro}
In modelling problems, prior knowledge about the function shape, like monotonicity, is often available. However, especially in complex, multidimensional models, such prior knowledge might not be available, yet can clearly be seen from the data. In these cases it would be beneficial to find such shape information and use it to increase model performance. Usually this is of interest especially for non-parametric function estimation, since these methods are the natural choice for complex, multidimensional data when the shape of the function is uncertain.  

Gaussian Processes (GPs) are a flexible, tractable and widely used non-parametric function estimation method, where priors are defined over latent functions \citep{rasmussen2006gaussian}. One advantage for GPs in this context is the fact that, using derivative information with GPs is easy as a derivative of a GP is also a GP. Furthermore, utilizing derivative information is substantially known to improve estimation performance of nonlinear dynamic systems \citep{solak2003derivative}. For GPs, many methods for using prior knowledge about the function shape have been presented. \cite{riihimaki2010gaussian} proposed a method, where virtual derivative observations are assumed at a finite number of locations to enforce monotonicity. They approximate the joint posterior with expectation propagation (EP) \citep{minka2001family} and maximize the resulting approximate likelihood to find point estimates for the hyperparameters.

In this paper we apply these methods to input variables without known monotonicities and use model comparison methods to automatically detect monotonic input dimensions. The added value of automatically detected monotonicities can best be seen near the edges of the data where the data is normally sparser and the model does not fit that well. In addition to the edges, the method is expected to be useful when extrapolating outside the input data.

This paper is structured as follows. Section~\ref{sec:background} briefly presents the background of GPs with derivative observations and the theory behind virtual derivative observations. Section~\ref{sec:amd} presents the proposed framework for automatically selecting monotonic input dimensions. Section \ref{sec:experiments} presents two experiments that were performed with the proposed method on synthetic and real data. Finally, Section~\ref{sec:discussion} concludes the paper and identifies possible topics for further research.




\section{BACKGROUND}
\label{sec:background}

In this section we briefly revise the theoretical background of Gaussian processes and motivate the problem behind this paper. 

\subsection{Gaussian processes with derivative observations}
Let $\mathbf{x}\in \mathbb{R}^{ 1 \times D}$ denote a $D$-dimensional input vector. The GP prior is directly specified on the latent function $f(\mathbf{x})$ and the prior assumptions are encoded in the covariance function $k(\mathbf{x}_1, \mathbf{x}_2)$, which specifies the covariance of two latent function values $f(\mathbf{x}_1)$ and $f(\mathbf{x}_2)$. We assume a zero mean Gaussian process prior
\begin{equation} \label{eq:gp}
p(\mathbf{f} | \mathbf{X}) = N(\mathbf{f} | \mathbf{0}, \mathbf{K}),
\end{equation}
where $\mathbf{K}$ is a covariance matrix between $N$ latent values $\mathbf{f} \in \mathbb{R}^{N \times 1}$ at the training inputs $\mathbf{X} = \left( \mathbf{x}_1 \tr, \ldots, \mathbf{x}_n \tr  \right)$, s.t. $\mathbf{K}_{ij} = k(\mathbf{x}_i, \mathbf{x}_j)$.

In the regression case, $N$ noisy observations $\mathbf{y} \in \mathbb{R}^{N \times 1}$ and latent function values are assumed to have Gaussian relationship. The joint distribution of the observations $\mathbf{y}$ and $M$ latent values $\mathbf{f}_* \in \mathbb{R}^{M \times 1}$ at the test inputs $\mathbf{X}_* \in \mathbb{R}^{M \times D}$ is
\begin{equation} \label{eq:gj}
\left[\begin{matrix}\mathbf{y} \\ \mathbf{f}_*\end{matrix} \right]  \sim N \left( \mathbf{0}, \left[\begin{matrix} \mathbf{K} + \sigma^2 \mathbf{I} & \mathbf{K}_* \tr \\ \mathbf{K}_* & \mathbf{K}_{**}  \end{matrix}  \right]  \right),
\end{equation}
with $\sigma^2$ being the noise variance, $\mathbf{I} \in \mathbb{R}^{N \times N}$ identity matrix, $\mathbf{K}_* \in \mathbb{R}^{N \times M}$ the covariance between the latent values at the training and test inputs and $\mathbf{K}_{**} \in \mathbb{R}^{M \times M}$ the covariance matrix of the latent values at the test inputs. With the Gaussian conditioning rule, the predictive distribution becomes
\begin{align*}
\mathbf{f}_* | \mathbf{y}& \sim N( \mathbf{f}_* | \boldsymbol{\mu}_*, \boldsymbol{\Sigma}_*), \\
\boldsymbol{\mu}_* & = \mathbf{K}_* (K + \sigma^2 \mathbf{I})^{-1} \mathbf{y}, \\
\boldsymbol{\Sigma}_* & = \mathbf{K}_{**} - \mathbf{K}_* (\mathbf{K} + \sigma^2 \mathbf{I})^{-1} \mathbf{K}_* \tr.
\end{align*}

As the differentiation is a linear operator, the partial derivative of a Gaussian process remains a Gaussian process \citep{solak2003derivative, rasmussen2006gaussian}. Thus, using partial derivative values for prediction and making predictions about the partial derivatives at a given point is easy. As
\begin{align*}
\cov{ \frac{\partial f^{(i)}}{\partial x^{(i)}_g}, \; f^{(j)} } = \frac{\partial }{\partial x^{(i)}_g}\; \cov{ f^{(i)}, \; f^{(j)}}, \\
\cov{ \frac{\partial f^{(i)}}{\partial x^{(i)}_g}, \; \frac{\partial f^{(j)}}{\partial x^{(j)}_h} } = \frac{\partial^2 }{\partial x^{(i)}_g \partial x^{(j)}_h  }\; \cov{ f^{(i)}, \; f^{(j)}}
\end{align*}
covariance matrices of Equations \eqref{eq:gp} and \eqref{eq:gj} can be extended correspondingly depending on whether the values of $\mathbf{y}$ are partial derivative observations, or partial derivative predictions are wanted on test points $\mathbf{X}_*$. 


\subsubsection{Expressing monotonicity with virtual observations}

Since GP is a nonparametric model, there is no single parameter from which the monotonicity could be checked. Because of this, the monotonicity needs to be confirmed from the gradient. However, since GPs are assumed to be smooth, gradient evaluations need to be done only in finite number of points. The degree of the smoothness and interval of gradient evaluations depends on the used covariance function and prior distribution of its hyperparameters. The method presented in this section for expressing monotonicity, is originally presented by \cite{riihimaki2010gaussian}.

Let $m^{(i)}_{d_i}$ be the partial derivative information at $\mathbf{x}^{(i)}$ in the dimension $d_i$. To express monotonicity $m^{(i)}_{d_i}$, probit likelihood is assumed for the partial derivative observations
\begin{equation*}
p\left( m^{(i)}_{d_i} \given \frac{\partial f^{(i)}}{\partial x_{d_i}^{(i)}} \right) = \boldsymbol{\Phi} \left( \frac{\partial f^{(i)}}{\partial x_{d_i}^{(i)}} \frac{1}{\nu} \right),
\end{equation*}
where
\begin{equation*}
\boldsymbol{\Phi}(z) = \int_{- \infty} ^{z} N(t|0,1) dt.
\end{equation*}
Here $\nu$ is a control parameter for the steepness of the likelihood. As $\nu \rightarrow 0$, probit likelihood approaches the step function and the likelihood penalizes more for erroneous derivative information. Let $\bf{m}$ be a set of $M$ partial derivative observations at $\bf{X}_m$. Assuming conditional independence given the latent derivative values, likelihood becomes
\begin{equation*}
p(\mathbf{m} \left. \given \right. \mathbf{ f'_{X_m} }) = {\prod_{i=1}^M}  p \left( m^{(i)}_{d_i} \given \frac{\partial f^{(i)}}{\partial x_{d_i}^{(i)}} \right).
\end{equation*}
In this paper the model selection is done with model comparison. Because of this, the tested monotonicity information can be assumed to be certain. Model selection method takes care that no false assumptions are led to the final model. Thus probit likelihood can be very steep and $\nu = 10^{-6}$ is used. Furthermore, since for steep probit likelihood, the likelihood values are close to each other for $ z>0$ and for $ z<0$, the assumed numeric value does not matter and values $m^{(i)}_{d_i} = \pm 1$ are used $\forall i = i,\ldots,M$ to express increasing/decreasing functions

Given a set of points $\mathbf{X}_m$, where the function is known to be monotonic, and set of points $\bf{X}$, the prior for latent values $\bf{f_X}$ and wanted latent value derivatives $\bf{f'_{X_m}}$ becomes
\begin{equation*}
\begin{aligned}
&p\left(\left[ \begin{matrix} \mathbf{f_X} \\ \mathbf{f'_{X_m}} \end{matrix} \right]  \given \left[\begin{matrix} \mathbf{X} \\ \mathbf{X_m}\end{matrix} \right]\right) = N \left( \mathbf{f}_{\text{joint}} \given \mathbf{0}, \mathbf{K}_{\text{joint}}  \right)  \\ &  =    N\left(\left[ \begin{matrix} \mathbf{f_X} \\ \mathbf{f'_{X_m}}\end{matrix} \right] \given \mathbf{0}, \left[ \begin{matrix} \mathbf{K_{f_X, f_X}} & \mathbf{K_{f_X, f'_{X_m}}}  \\ \mathbf{K_{ f'_{X_m},f_X}} & \mathbf{K_{ f'_{X_m}, f'_{X_m}}} \end{matrix} \right]\right).
\end{aligned}
\end{equation*}

\begin{figure}[!t]
\centering
\includegraphics{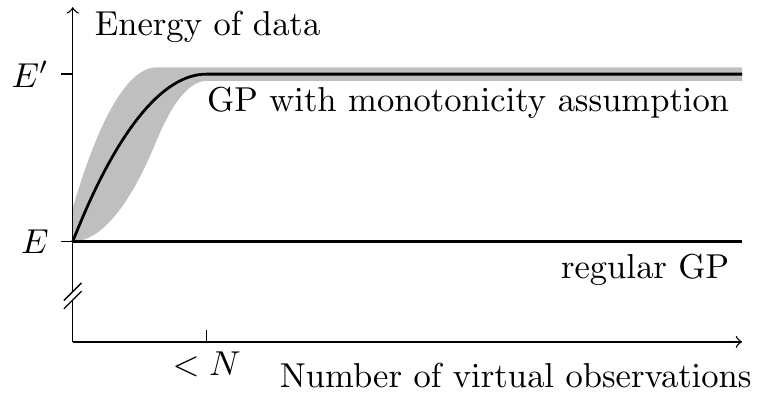}
\caption{Energy function of data as a function of virtual observations for GPs for one dimensional data with and without monotonicity assumption. The gray area around the fitted GP with monotonicity assumption describes the uncertainty in the energy value that is caused by the randomness of the virtual derivative observation locations, randomness of the data and differences in the underlying true function.  \label{fig:EF}}
\end{figure}

The joint posterior for latent values and latent value derivatives can be obtained from Bayes' rule
\begin{equation}
\begin{aligned}
&p(\mathbf{f_X}, \mathbf{f'_{X_m}}|\mathbf{y}, \mathbf{m}, \mathbf{X}, \mathbf{X_m}) \\ & = \frac{p(\mathbf{f_X}, \mathbf{f'_{X_m}}| \mathbf{X}, \mathbf{X_m})p(\mathbf{y}|\mathbf{f_X})p(\mathbf{m}|\mathbf{f'_{X_m}})}{Z},
\end{aligned}
\label{eq:posterior}
\end{equation}
where $Z$ is the normalizing term, or the marginal likelihood
\begin{equation*}
\begin{aligned}
&Z= p(\mathbf{y},\mathbf{m} | \mathbf{X}, \mathbf{X_m}) \\ &=\int\! p(\mathbf{f_X}, \mathbf{f'_{X_m}}| \mathbf{X}, \mathbf{X_m})p(\mathbf{y}|\mathbf{f_X})p(\mathbf{m}|\mathbf{f'_{X_m}}) d \mathbf{f_X} d \mathbf{f'_{X_m}}.
\end{aligned}
\end{equation*}
Since $p(\bf{m}|\bf{f'_{X_m}})$ is not Gaussian, the full posterior is analytically intractable and some approximation method must be used. In a paper by \cite{riihimaki2010gaussian}, the posterior is approximated by expectation propagation (EP) algorithm and in a paper by \cite{wang2016estimating} by Gibbs sampling. We use EP as its approximation properties are proven to be good, it is computationally faster and computing the marginal likelihood is easy. The last property is especially useful for model comparison methods used in this paper and explained later.

In EP, the posterior distribution in Equation \eqref{eq:posterior} is approximated by
\begin{equation}
\begin{aligned}
&q\left( \mathbf{f_X}, \mathbf{f'_{X_m} | \mathbf{y}, \mathbf{m}, \mathbf{X}, \mathbf{X_m}}\right) \\ & = \frac{ p(\mathbf{f_X}, \mathbf{f'_{X_m}}| \mathbf{X}, \mathbf{X_m})p(\mathbf{y}|\mathbf{f_X}) \prod_{i=1}^M t_i( \tilde{Z}_i , \tilde{\mu}_i, \tilde{\sigma}_i ^2) }{Z_{EP} },
\end{aligned}
\label{eq:EPposterior} 
\end{equation}
where $t_i( \tilde{Z}_i , \tilde{\mu}_i, \tilde{\sigma}_i ^2) = \tilde{Z}_i N(f'_{X_{m}i} | \tilde{\mu}_i, \tilde{ \sigma}^2 _i  )$ are the local likelihood approximations. The approximation for the marginal likelihood becomes
\begin{equation}
\begin{aligned}
& Z_{EP} = q(\mathbf{y},\mathbf{m} | \mathbf{X}, \mathbf{X_m}) = \int \!p(\mathbf{f_X}, \mathbf{f'_{X_m}}| \mathbf{X}, \mathbf{X_m})\\& \times p(\mathbf{y}|\mathbf{f_X}) \prod_{i=1}^M t_i( \tilde{Z}_i , \tilde{\mu}_i, \tilde{\sigma}_i ^2) d \mathbf{f_X} \mathbf{f_{X_m}}.
\end{aligned}
\label{eq:EPmarg}
\end{equation}
 Since everything in Equation \eqref{eq:EPposterior}, is Gaussian, the posterior approximation simplifies to
\begin{equation*}
\begin{aligned}
&q\left( \mathbf{f_X}, \mathbf{f'_{X_m} | \mathbf{y}, \mathbf{m}, \mathbf{X}, \mathbf{X_m}}\right) \\ & = N\left(\mathbf{f}_{\text{joint}} \given \boldsymbol{\Sigma} \tilde{\boldsymbol{\Sigma}}^{-1}_{\text{joint}} \boldsymbol{\mu}, \left(\boldsymbol{K}_{\text{joint}}^{-1} + \tilde{\boldsymbol{\Sigma}}_{\text{joint}}^{-1}\right)^{-1}\right)
\end{aligned}
\end{equation*}
\begin{align*}
\text{with } \tilde{\boldsymbol{\mu}}_{joint} = \left[ \begin{matrix}\mathbf{y} \\ \tilde{\boldsymbol{\mu}}\end{matrix} \right],\;\text{and } \tilde{\boldsymbol{\Sigma}}_{joint} = \left[ \begin{matrix} \sigma^2 \mathbf{I} &  \mathbf{0} \\ \mathbf{0} & \tilde{\boldsymbol{\Sigma}} \end{matrix} \right].
\end{align*}
Here $\tilde{\boldsymbol{\mu}}$ consists of site means $\tilde{\mu}_i$ and $\tilde{\boldsymbol{\Sigma}}$ is a diagonal matrix, with site variances $\tilde{\sigma}^2 _i$. These parameters can be computed with the standard binary classification EP algorithm, which is described for example in chapter 3.6 of \cite{rasmussen2006gaussian}.

Model comparison is often done with the energy function, or negative log marginal likelihood, of the data
\begin{equation*}
E(\mathbf{y}| \mathbf{X}) = - \log p(\mathbf{y}| \mathbf{X}) \label{eq:energy}
\end{equation*}
Given a clearly monotonic function, adding virtual derivative observations $\mathbf{m}$ with probit likelihood and $\nu$ close to zero, will not crucially affect the probability, or energy function, of the data
\begin{equation}
\begin{aligned}
& E(\mathbf{y},\mathbf{m}| \mathbf{X}, \mathbf{X_m}) = - \log p(\mathbf{y},\mathbf{m} | \mathbf{X}, \mathbf{X_m})  \\ & = -\log \left( p(\mathbf{y} | \mathbf{X})\overbrace{p(\mathbf{m} | \mathbf{y}, \mathbf{X}, \mathbf{X_m})}^{\approx 1} \right) \approx E(\mathbf{y}| \mathbf{X}).
\end{aligned} 
\label{eq:pconcept}
\end{equation}
Assuming that virtual derivative observations have been divided to two sets $\{ \mathbf{X}_{\mathbf{m}^1}, \mathbf{m}^1 \}$ and $\{ \mathbf{X}_{\mathbf{m}^1}, \mathbf{m}^2 \}$, where only one is enough to convince that the model is monotonic. In this case
\begin{equation*}
\begin{aligned}
& p(\mathbf{y},\mathbf{m}^1, \mathbf{m}^2 | \mathbf{X}, \mathbf{X}_{\mathbf{m}^1}, \mathbf{X}_{\mathbf{m}^2})  \\ &= p(\mathbf{y}, \mathbf{m}^1 | \mathbf{X}, \mathbf{X}_{\mathbf{m}^1})\overbrace{p(\mathbf{m}^2 | \mathbf{y},\mathbf{m}^1, \mathbf{X}, \mathbf{X}_{\mathbf{m}^1}, \mathbf{X}_{\mathbf{m}^2})}^{\approx 1} \\
& \approx p(\mathbf{y}, \mathbf{m}^1 | \mathbf{X}, \mathbf{X}_{\mathbf{m}^1}).
\end{aligned} 
\end{equation*}
So, if enough virtual observations are added, adding more of them does not affect the energy. Our test have shown that fairly few virtual observations are enough to approximate the monotonicity constraint. For example, in one dimensional case, despite the data, the number of virtual observations needed to stabilize energy function is a lot less than the number of data points $N$. This concept is visualised in Figure \ref{fig:EF}.

In Figure \ref{fig:EF}, the difference between final energy values of GP with and without virtual derivative observations, $E$ and $E'$, depends merely on how well the data supports the virtual observations. If the data and the virtual derivative observations agree, $E \approx E'$, but if the support is weak, there is no theoretical upper limit for $E'$. 

The locations of the derivative observations can be selected in many ways \citep{riihimaki2010gaussian}. In low dimensions, the observations can be placed on a grid. The drawback of this approach is that as the dimensionality grows, the number of observations needed increases exponentially. Another strategy is to select virtual points randomly from some distribution. This distribution can be uniform or then try to mimic the distribution of the samples. More elaborate methods have also been studied. As the computational scaling is $\mathcal{O}((N+M)^3)$, it is often desirable that the number of virtual observations is as small as possible. In this case, the locations can be chosen iteratively so that one point is added at a time and the next point is selected from the posterior distribution according to some heuristic. 

\section{AUTOMATIC MONOTONICITY DETECTION FRAMEWORK FOR GAUSSIAN PROCESSES}
\label{sec:amd}

\begin{algorithm}[b!]
\caption{Algorithm for automatic monotonicity detection (AMD) \label{alg:amd}}
\label{AMD}
\begin{algorithmic}[1]
\Function{AMD}{$\mathbf{X} \in  \mathbb{R}^{N \times D}, \mathbf{y} \in \mathbb{R}^{N \times 1}, p_1,p_2$}
\State  $E_{\text{bl1}} \leftarrow E_{\text{EP}}(\mathbf{y}| \mathbf{X})- p_1 \times \frac{N}{2} \log 2 \pi$ 
\State  $E_{\text{bl2}} \leftarrow E_{\text{EP}}(\mathbf{y}| \mathbf{X})- p_2 \times \frac{N}{2} \log 2 \pi$ 
\State $\mathbf{d} \leftarrow \mathbf{0}_{D \times 1}$
\For{ $i \in [1, D]$ }
\State Select $\mathbf{X_{m^{(i)}}} \in \mathbb{R}^{M \times D}$ 
\State $\mathbf{m^{(i)}} \leftarrow \mathbf{1}_{M \times 1} $
\State  $ E_{\text{mon+}}^{(i)} \leftarrow E_{\text{EP}}(\mathbf{y}, \mathbf{m^{(i)}}|\mathbf{X}, \mathbf{X_{m^{(i)}}})$
\State  $ E_{\text{mon-}}^{(i)} \leftarrow E_{\text{EP}}(\mathbf{y}, -\mathbf{m^{(i)}}|\mathbf{X}, \mathbf{X_{m^{(i)}}})$
\If{$E_{\text{mon+}}^{(i)} \leq E_{\text{bl1}}$ and $E_{\text{mon-}}^{(i)} > E_{\text{bl2}}$}
\State $\mathbf{d}_i  \leftarrow +1 $
\ElsIf{$E_{\text{mon-}}^{(i)}\! \leq \! E_{\text{bl1}}$ and $E_{\text{mon+}}^{(i)}\! > \! E_{\text{bl2}}$}
\State $\mathbf{d}_i  \leftarrow -1 $
\EndIf
\EndFor
\State \Return $\mathbf{d}$
\EndFunction
\end{algorithmic}
\end{algorithm}


This section presents our proposed method for automatically finding monotonic input dimensions from data. The only requirement of the proposed method is that a GP with Gaussian likelihood and covariance function with computable first and second derivatives can be fitted to the data. The methodology could be extended also to non-Gaussian likelihoods, but we do not focus on them in this paper.

The simple idea of energy comparisons of Equation \eqref{eq:comparison} forms the basis of our automatic monotonicity detection (AMD) framework for GPs. Given the data, different monotonicity combinations can be compared using their energy functions. However, as probit likelihood makes analytical solution for marginal likelihood $p(\mathbf{y},\mathbf{m} | \mathbf{X}, \mathbf{X_m})$ intractable, it has to be approximated.

In our method, the energy of the data is approximated using EP. Since everything in Equation \eqref{eq:EPmarg} is Gaussian (and integral of Gaussian is Gaussian), the marginal posterior simplifies to
\begin{equation*}
Z_{\text{EP}} = Z_{\text{joint}}^{-1} \prod_{i=1}^M \tilde{Z}_i,
\end{equation*}
where
\begin{align*}
Z_{joint}^{-1} = (2 \pi)^{-\frac{N+M}{2}} | \mathbf{K}_{joint} + \tilde{\boldsymbol{\Sigma}}_{joint}|^{-\frac{1}{2}} \\ \times \exp \left( -\frac{1}{2}\tilde{\boldsymbol{\mu}}_{joint}\tr \left( \mathbf{K}_{joint} + \tilde{\boldsymbol{\Sigma}}_{joint} \right)^{-1} \tilde{\boldsymbol{\mu}}_{joint} \right),
\end{align*}
and the energy function of the data can be expressed as
\begin{equation}
\begin{aligned}
&- \log q(\mathbf{y},\mathbf{m} | \mathbf{X}, \mathbf{X_m}) =  - \frac{1}{2} \log |\mathbf{K}_{joint} + \tilde{\boldsymbol{\Sigma}}_{joint}| \\ & - \frac{1}{2} \tilde{\boldsymbol{\mu}}_{joint}\tr \left( \mathbf{K}_{joint} + \tilde{\boldsymbol{\Sigma}}_{joint} \right)^{-1} \tilde{\boldsymbol{\mu}}_{joint} \\ &  + \frac{1}{2} \sum_{i=1}^M \frac{(\mu_{-i}-\tilde{\mu}_i)^2}{(\sigma_{-1}^2 + \tilde{\sigma}_i ^2)}  + \sum_{i=1}^M \log \Phi \left( \frac{\mu_{-i}}{ \sqrt{\nu^2 + \sigma_{-i}^2}} \right)\\ & +\frac{1}{2} \sum_{i=1}^M \log \left( \sigma_{-i}^2 + \tilde{\sigma}_{-i}^2 \right).
\end{aligned} \label{eq:EPenergy}
\end{equation}

Since EP is an approximation, Equation \eqref{eq:pconcept} does not fully hold. However, our empirical tests have shown that despite the dimensionality or number of virtual observations for clearly monotonic functions with Gaussian likelihood
\begin{equation*}
E_{EP}(\mathbf{y}, \mathbf{m}| \mathbf{X}, \mathbf{X_m}) \approx E_{EP}(\mathbf{y}|\mathbf{X}) -\frac{N}{2} \log 2 \pi,
\end{equation*}
where the added term $-\frac{N}{2} \log 2 \pi$ is a result of approximations done in Equation \eqref{eq:EPenergy}.

For each dimension, there are three possibilities for the monotonicity. These are increasing, decreasing or non-monotonic with respect to that dimension. Because of this, in multidimensional case, there are $3^D$ possible combinations. However, computing the probability of all of them is not necessary in order to find the correct monotonicity combination. Since Equation \eqref{eq:pconcept} also holds if $\mathbf{m}$ contains virtual observations of a single dimension, it is enough to individually test monotonicity of each input dimension. Because of this, from this on let $\mathbf{m}^{(i)}$ denote a monotonicity assumption of dimension $i$ .

\begin{figure*}[t!]
    \centering
    \includegraphics{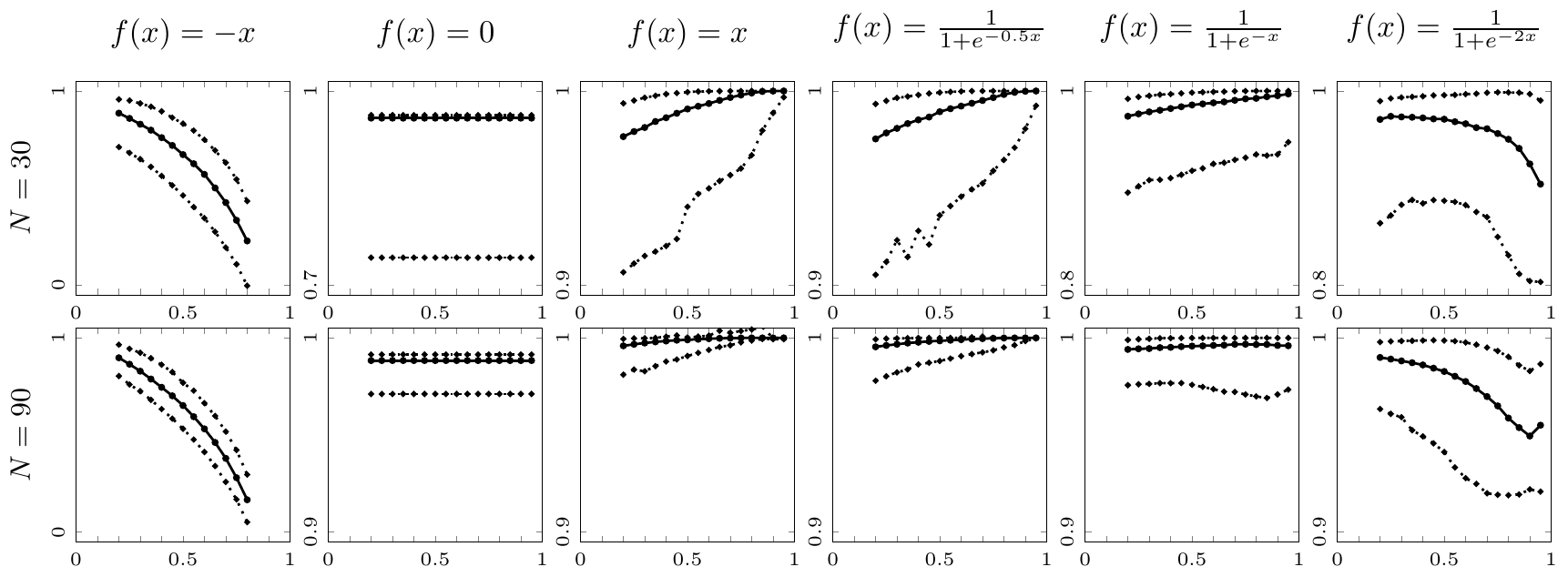}
\caption{Median and 90\% central interval for the upper limit of $p_1$ from Equation \eqref{eq:comparison} as a function of variance of the signal. Columns define different functions and rows define number of training samples. As the data is normalized, variance of the signal is same as proportion of the variance explained by the signal, $\frac{\sigma^2_{\text{signal}}}{\sigma^2_{\text{signal}} + \sigma^2_{\text{noise}}}$. Notice that each plot has different scale for y-axis. From the Figure it can be seen for example that for $f(x)=\frac{1}{1-e^{-x}}$, $\sigma^2_{\text{signal}}=0.7$ and $N=30$, value $p_1=0.9904$ would result to positive monotonicity classification in half of the cases.}
\label{fig:proportion}
\end{figure*}

Since there are different uses of models, the degree of how much data has to agree with the assumed shape constraint varies. Because of this, the underlying function is assumed to be monotonic with monotonicity assumption $\mathbf{m}^{(i)}$ if 
\begin{equation}
\begin{aligned}
&E_{EP}(\mathbf{y}, \mathbf{m}^{(i)}| \mathbf{X}, \mathbf{X_{m^{(i)}}})\! \leq E_{EP}(\mathbf{y}|\mathbf{X}) - p_1 \!\!\times\!\! \frac{N}{2} \log 2 \pi \\ &\text{and} \\
&E_{EP}(\mathbf{y}, -\mathbf{m^{(i)}}| \mathbf{X}, \mathbf{X_{m^{(i)}}})\! > E_{EP}(\mathbf{y}|\mathbf{X}) - p_2\!\!\times \!\! \frac{N}{2} \log 2 \pi .
\end{aligned}
\label{eq:comparison}
\end{equation} 
In above $- \mathbf{m^{(i)}}$ if opposite monotonicity assumption to $\mathbf{m}$ and $\{p_1, p_2 \} \in ( -\infty, 1], p_1 > p_2 $ are predefined confidence coefficients that define how easily different monotonicity assumption is accepted. Intuitively, $p_1$ defines how close the energy of assumed monotonicity $\mathbf{m^{(i)}}$ has to be to the regular GPs energy value and $p_2$ defines how far we want the energy of opposite monotonicity assumption to be from the regular GP's energy. For $p_1$ close to one and $p_2$ having small values, the monotonicity assumption $\mathbf{m}$ requires lots of evidence from the data to be accepted. The relationship of $p_1$ and $p_2$ are further analysed in Section \ref{sec:real}. The whole automatic monotonicity detection (AMD) algorithm is described in the Algorithm \ref{alg:amd}.

\section{EXPERIMENTS}
\label{sec:experiments}
\subsection{Test cases with synthetic data \label{sec:synthetic}}

To prove that the concept described in the Section \ref{sec:amd} works, it is tested with different one dimensional synthetic data sets that mimic different degrees of monotonicity and noisiness. The synthetic data were created as follows. As real data often is not distributed equally in the input space, the input $\mathbf{X}$ of our synthetic data is drawn from normal distribution. The observations are drawn from two function families
\begin{align}
&f(x) = ax,\;\; a\in\{-1,0,1\} \label{eq:bl}\\
&f(x) = \frac{1}{1+e^{-ax}},\;\; a \in \mathbb{R}_{> 0}. \label{eq:sigmoid}
\end{align}
Linear function (Equation \eqref{eq:bl}) is used as a baseline for monotonicities of different order. When $a=0$, the function is an example of a non-monotonic function that can easily be confused as monotonic one. When $a=-1$, the function is used as a baseline for a function that is clearly monotonic in the wrong direction. Finally, when $a=1$, the function is a model example of a monotonic function. Sigmoid function (Equation \eqref{eq:sigmoid}) is used with different values for $a$ to resemble different kind of monotonic functions. 

To mimic different noise levels, normally distributed noise with different variances is added to the data to adjust signal to noise ratio $\frac{\sigma^2_{\text{signal}}}{\sigma^2_{\text{noise}} + \sigma^2_{\text{signal}}}$. After the noise is added, the data is normalized to zero mean and unit variance. However, for $f(x)=0$, the noise variance is kept constant, since the variance of the signal is zero. 

\begin{figure*}[t!]
    \centering
    \includegraphics{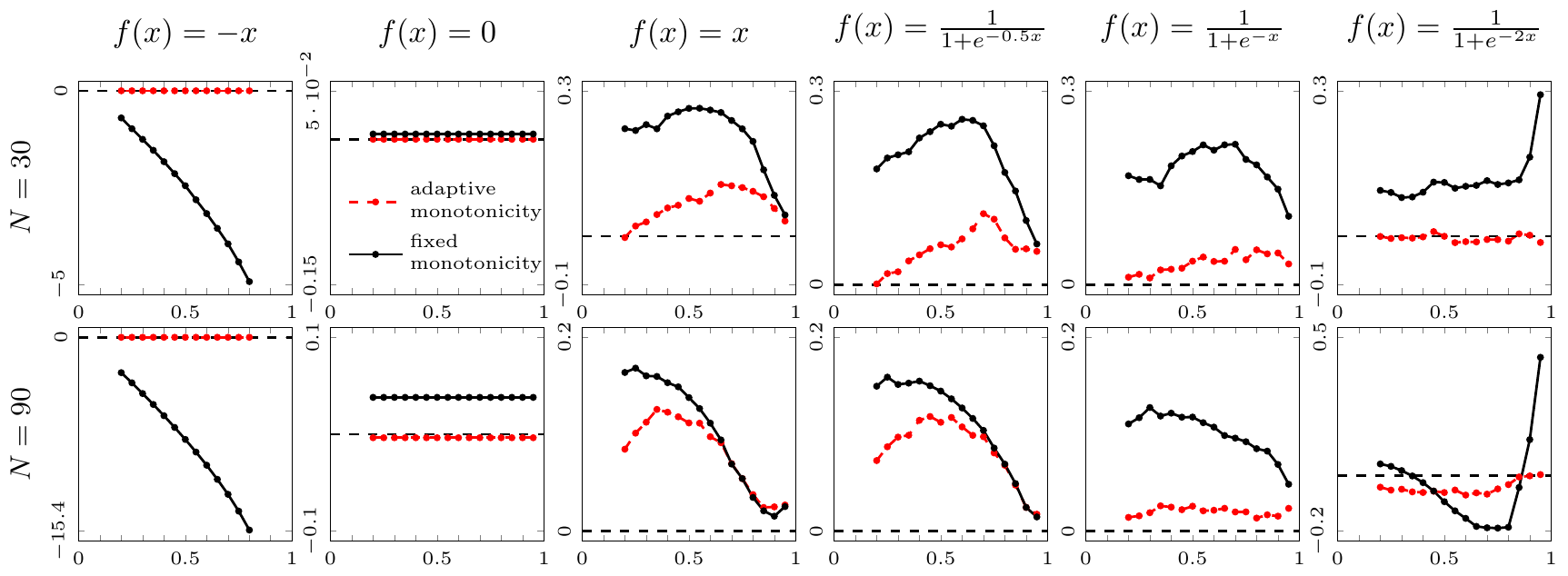}
\caption{Change in lppd-value near the edges of the test data as a function of proportion of variance explained by the signal when switching from regular GP to monotonic GP and AMD. The dashed line is drawn for reference to where change in lppd is zero. Because of this, data point above dashed line means increased performance. Columns define different functions and rows define number of training samples. $p_1=0.99$ is used for AMD.}
\label{fig:lpd} 
\end{figure*}

To find out possible values for $p_1$ from Equation \eqref{eq:comparison}, upper limit with monotonicity assumption $\mathbf{m}^{(\text{i})}$ becomes 
\begin{equation*}
p_1 = \frac{ E_{\text{EP}}(\mathbf{y} \left. \given \right. \mathbf{X})-  E_{\text{EP}} ( \mathbf{y}, \mathbf{m}^{(\text{i})} \left. \given \right. \mathbf{X}, \mathbf{X}_{\mathbf{m}^{(\text{i})}} ) }{\frac{N}{2} \log 2 \pi}.
\end{equation*}
To find out good values for $p_1$, the data is simulated 200 times and upper limit is computed. This value as a function of the proportion of variance explained by the signal for functions of Equations \eqref{eq:bl} and \eqref{eq:sigmoid} are visualised in Figure \ref{fig:proportion} for training data sizes of $N=\{30,90\}$. Number of virtual derivative observations is kept at $M=\frac{N}{3}$ in all examples and the observations are placed to an uniform grid that covers the input space. The used GP has Gaussian likelihood and squared exponential covariance function. Squared exponential covariance function and its required derivatives are presented, for example, in \citep{riihimaki2010gaussian}.

The results of Figure \ref{fig:proportion} show that the upper limit for $p_1$ is very close to 1 for all monotonic functions. This even holds with relatively small data set size and high noise levels. However, data generated from functions $f(x)=0$ and $f(x) = \frac{1}{1+e^{-2x}}$ cause problems. For zero function the reason is clear, since adding the virtual derivative observations compromises with the data only a little and thus upper limit gets so high values. However this is not a problem for Algorithm \ref{alg:amd}, since the comparison of energies would also be made with virtual derivative observation of negative monotonicity and with $p_2$ set to low value, false monotonicity assumption would not be made. For sigmoid function with $a=2$, the results are more concerning especially with less training samples. The only way to increase the probability of correct classification in this case would be to make values of $p_1$ smaller and simultaneously accept the risk of falsely classifying some non-monotonic function as monotonic.

To see the change in the prediction accuracy of the data caused by the positive monotonicity detection, log pointwise predictive density (lppd) of data $\mathbf{X} \in \mathbb{R}^{L \times D}$, $\mathbf{y} \in \mathbb{R}^{L \times 1}$ is used
\begin{equation}
\text{lppd} = \sum_{i=1}^L \log \int p(y_i | f) p_{\text{post}}(f| x_i) d f.
\end{equation}
As discussed in the Section \ref{sec:intro}, the method is expected to be most beneficial at the borders of the data, where there normally is less data. Because of this, test data points and observations are generated from the same distribution as the original training data and $20 \%$ of the most outermost values and corresponding observations are selected. Lppd is computed for each function and data set size for assumed positive monotonicity, no monotonicity and simplified version of AMD algorithm, where the opposite monotonicity assumption is not computed. To make comparison to automatic monotonicity detection easy, difference in lppd between automatic monotonicity detection and non monotonic GPs, and increasing GP and non monotonic GPs are plotted. Value $p_1=0.99$ is used in all cases and the results are averaged from 200 evaluations. The results are illustrated in Figure \ref{fig:lpd}.

\begin{figure*}[t!]
\includegraphics{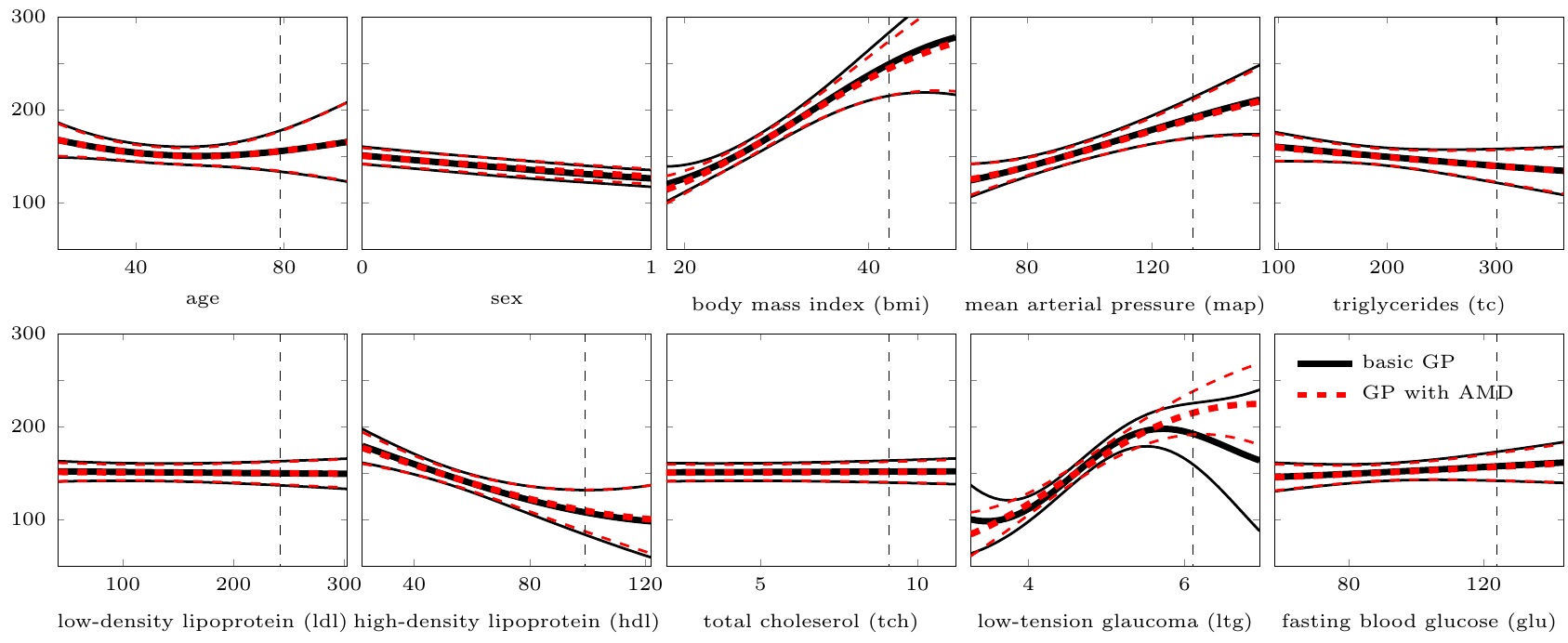}
\caption{Target value, a measure of diabetes progression one year after baseline, as a function of single predictive values while others are kept at the median of dataset. Regular black lines correspond to regular GP mean and 90 \% posterior central interval. Red dashed lines correspond to AMD GPs mean and standard deviation when body mass index and low-tension glaucoma are detected as increasing. Black dashed line corresponds to the largest value of covariate. Notice that the sex is a binary covariate but it has been modeled as continous. \label{fig:diabetes} }
\end{figure*}

The results of Figure \ref{fig:lpd} show that automatic positive monotonicity detection can improve the predictive accuracy of the test data when the monotonicity assumption is correct and the error made when monotonicity assumption is false, is very little. The latter property is bad, but the decrease in performance is very little and is overcome by the increased performance in case of correct assumption. However, same functions cause problems as with the classification accuracy. Similarly as before, if also negative monotonicity was tested, results for zero function would be better because of $p_2$. Weirdly, even though sigmoid function is monotonic with $a=2$, the correct virtual derivative observations make the lppd results worse than for regular GP. The reason for this has been explained in \citep{riihimaki2010gaussian}, where it was shown that for step function, use of EP and virtual derivative observations lead to too long lengthscale hyperparameter and thus the performance decreases. Now as $a=2$, sigmoid function is close to step function and the same problem occurs.

Even though limits for $p_2$ were not analysed, the upper limit for follows straight from the Equation \eqref{eq:comparison} and is 
\begin{equation*}
p_2 = \frac{ E_{\text{EP}}(\mathbf{y} \left. \given \right. \mathbf{X})-  E_{\text{EP}} ( \mathbf{y}, - \mathbf{m}^{(\text{i})} \left. \given \right. \mathbf{X}, \mathbf{X}_{\mathbf{m}^{(\text{i})}} ) }{\frac{N}{2} \log 2 \pi}.
\end{equation*} 
Function \eqref{eq:bl} with $a=-1$ is a decreasing function and in Figure \ref{fig:proportion} the upper limit for $p_1$ is computed for assumption of GP being increasing. Because of symmetry, this is same as if the lower limit for $p_2$ was computed with function \eqref{eq:bl} and $a=1$. From Figure \ref{fig:proportion} it can be seen that the value of $p_2$ decreases rapidly as the variance of the noise decreases.

\subsection{Multi dimensional test case with real data \label{sec:real}}

Diabetes data\footnote{diabetes data, available at: \url{http://web.stanford.edu/~hastie/Papers/LARS/diabetes.data}}, first introduced by \cite{efron2004least}, is used to test the automatic monotonicity detection method. The dataset consists of 10 baseline variables and one target value that were measured from 442 diabetes patients. The baseline variables consist of 4 physical variables of and 6 blood serum variables of a single patient. The target value is a measure of disease progression one year after baseline.

\begin{figure}[!t]
\centering
\includegraphics{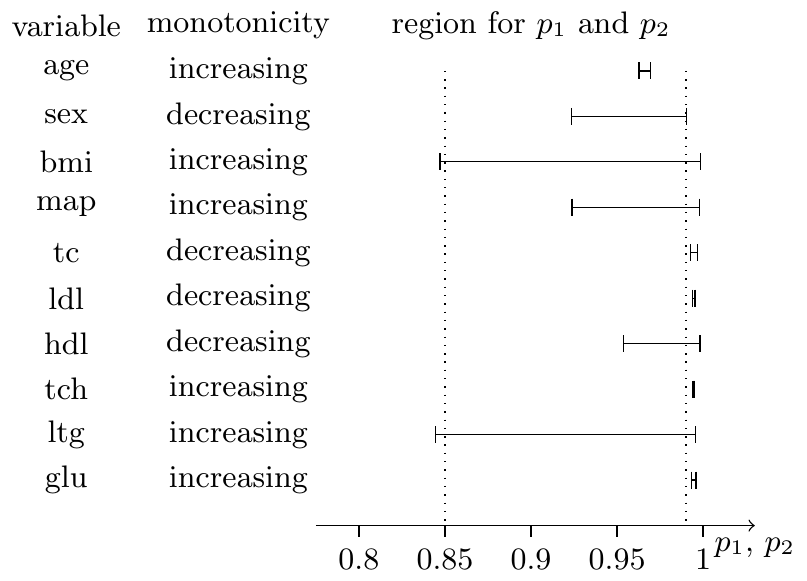}
\caption{Variable name, monotonicity direction and region where both $p_1$ and $p_2$ must be so that the variable would be detected as monotonic. For example if $p_1=0.99$ and $p_2=0.85$, bmi and ltg would be detected as monotonic. \label{fig:p}}
\end{figure}

Automatic monotonicity detection algorithm with $p_1=0.99$ and $p_2 = 0.85$ was used on the normalized data with GP having Gaussian likelihood and sum of squared exponential and linear function as its covariance function. The number of virtual observations was set to $\frac{N}{3}$. With this setting, the algorithm detected that two baseline variables, body mass index (bmi) and low-tension glaucoma (ltg), have positively monotonic relation to the target value. The results are illustrated in Figure \ref{fig:diabetes}. In addition to the actual results, it is useful to inspect how robust is the monotonicity detection to changes in the values of $p_1$ and $p_2$. Affect of changing $p_1$ and $p_2$ is illustrated in Figure \ref{fig:p}.

The results show that automatic monotonicity detection has significant change in some variables, especially near the borders of the data. As discussed in the Section \ref{sec:synthetic}, monotonicity detection increases the predictive accuracy especially near the borders. As it can be seen from Figure \ref{fig:diabetes}, where ltg being classified as  monotonic, visualisation methods of data can leave some aspects of the data unnoticed. It can clearly seen from the robustness visualization, Figure \ref{fig:p}, that data gives clear evidence of ltg being monotonic, but the visualization doesn't reveal the monotonicity that clearly. Same applies to other direction as well. The visualization gives evidence that also map is increasing, but the energy comparison does not give enough evidence for this.

For this data, two clearly detectable behaviours can be seen for the monotonicity detection of variables when $p_1$ and $p_2$ are changed. If the relation between variable and response is clearly monotonic, the monotonicity detection range for $p_1$ and $p_2$ is wide and the upper limit is very close to 1. This kind of behaviour can be detected from variables bmi and ltg. Also functions \eqref{eq:bl} with $a=\{-1, 1\}$ and \eqref{eq:sigmoid} with $a=\{0.5,1\}$ from the previous section would have similar kind of behaviour. If the effect of the variable is clearly non-monotonic and flat, the monotonicity detection range for $p_1$ and $p_2$ is very narrow. This kind of behaviour can be detected from variables age, tc, ldl, tch and glu. Also, the behaviour of zero function from the previous section would be really similar.

The lower limit for $p_2$ is surprisingly high when comparing it to the one dimensional test cases. The reason for this is the high dimensionality of the data. As the dimensionality increases, the changes in one dimension do not have that big effect in the other dimensions and thus the change in the posterior distribution is smaller. This can also be seen from the Figure \ref{fig:p}, where the affect of detected monotonicity can mostly be seen in the monotonic dimension.

\section{DISCUSSION}
\label{sec:discussion}
The proposed method has proved to be able to detect monotonic dimensions from different one-dimensional synthetic data and one multi-dimensional real life data set. Furthermore, the method can be expected to be usable in other data sets as well. The proposed algorithm can also be adjusted to detect monotonicities of different strictness by altering the values of $p_1$ and $p_2$.

Even though the proposed method has proved to work, there is no clear connection between the adjustable parameters and the strength of the monotonicity assumption. To solve this problem, reversible jump MCMC for sampling between different monotonicity combinations could be tried \citep{green1995reversible}. With this method, the results might be better as it would give probability for different monotonicities in all dimensions. However, even though the problem of the interpretability of the result would be solved, jumping MCMC causes problems when the parameter space grows. 

Another promising possible research topic would be to include interactivity to the monotonicity detection. After the proposed method finds possibly monotonic variables, it could ask from a data domain expert, whether or not the found monotonicity is probable. This kind of interactivity would be especially useful with high dimensional data.

Even though selection of the number of the virtual observations, $M$, was not the main emphasis of the proposed method, it would still be interesting to further study how small can $M$ be without making the results worse.

The proposed method was only analyzed for Gaussian likelihood. However, it would be possible to extend the method to other likelihoods as well. For example, comparing and analysing the energy of the data when the model uses probit likelihood, would make it possible to use the proposed method also with classification problems.

\subsubsection*{References}
\renewcommand{\section}[2]{}%
\bibliographystyle{apalike}
\bibliography{monodet.bbl}

\end{document}